\documentclass[aps,ams,amsmath,prx,twocolumn,superscriptaddress]{revtex4-1}
\usepackage{graphics}
\usepackage{amsmath}
\usepackage{amsfonts}
\usepackage{bm}
\usepackage{color}
\usepackage{bbm}
\usepackage{hyperref}
\def\bm#1{\mbox{\boldmath{$#1$}}}

\newcommand{\be}{\begin{equation}}
\newcommand{\ee}{\end{equation}}

\hypersetup{colorlinks=true,linkcolor=blue,anchorcolor=blue,citecolor=blue,filecolor=blue,urlcolor=blue,bookmarksnumbered=true,pdfview=FitB}

\newcommand{\beq}{\begin{equation}}
\newcommand{\eeq}{\end{equation}}
\newcommand{\beqa}{\begin{eqnarray}}
\newcommand{\eeqa}{\end{eqnarray}}
\newcommand{\bea}{\begin{eqnarray}}
\newcommand{\eea}{\end{eqnarray}}
\usepackage{feynmp-auto}
\usepackage{amssymb}
\usepackage{array}
\usepackage{amsmath}
\usepackage{graphicx}\usepackage{graphics}
\usepackage{dcolumn}
\usepackage{bm}\usepackage{varioref}
\DeclareMathOperator{\arctanh}{arctanh}
\DeclareMathOperator{\sgn}{sgn}

\newcommand{\St}{\mathrm{St}}

\begin{document}

\title{Conformal maps of viscous electron flow in the Gurzhi crossover}

\author{Songci Li}
\affiliation{Department of Physics, University of Wisconsin--Madison, Madison, Wisconsin 53706, USA}

\author{Maxim Khodas}
\affiliation{Racah Institute of Physics, Hebrew University of Jerusalem, Jerusalem 91904, Israel}

\author{Alex Levchenko}
\affiliation{Department of Physics, University of Wisconsin--Madison, Madison, Wisconsin 53706, USA}

\begin{abstract}
We investigate the impact of geometric constriction on the viscous flow of electron liquid through quantum point contacts. We provide analysis on the electric potential distribution given the setup of a slit configuration and use the method of conformal mapping to obtain analytical results. 
The potential profile can be tested and contrasted experimentally with the scanning tunneling potentiometry technique. We discuss intricate physics that underlies the Gurzhi effect, i.e., the enhancement of conductivity in the viscous flow, and compare results for different boundary conditions. In addition, we calculate the temperature dependence of the momentum relaxation time as a result of impurity assisted quasiballistic interference effects and discuss various correlational corrections that lead to the violation of Matthiessen's rule in the hydrodynamic regime. We caution that spatially inhomogeneous profiles of current in the Gurzhi crossover between Ohmic and Stokes flows might also appear in the nonhydrodynamic limit where nonlocality plays an important role. This conclusion is corroborated by calculation of dispersive conductivity in the weakly impure limit. 
\end{abstract}

\date{October 11, 2021}
 
\maketitle

\section{Introduction}

There has been a considerable amount of renewed interest in the study of electron transport effects in solid state systems that exhibit hydrodynamic behavior as originally formulated by Gurzhi \cite{Gurzhi}, see recent reviews \cite{NGMS,Lucas-Fong,ALJS} and references therein. Such a transport regime can occur in ultrapure and high mobility samples provided the momentum conserving electron-electron (ee) collisions are so frequent that the corresponding mean free path $l_{\text{ee}}$ is the smallest length scale as compared to other lengths at which momentum relaxation arises. Under these conditions the electron fluid equilibrates locally and is predicted to display several peculiar transport properties in confined geometries. The most prominent examples of electronic hydrodynamic behavior include current whirlpools concomitant with negative nonlocal voltages \cite{Torre,Levitov-1,Polini,Xie,Danz}, formation of a counterflow \cite{Titov}, viscous resistivity \cite{Hruska,AKS,Alekseev,LXA}, and superballistic conduction through the narrow constrictions \cite{Levitov-2} exceeding the limit given by the Sharvin formula \cite{Sharvin}. 

These transport phenomena in part have been observed experimentally \cite{Bandurin,Mackenzie,Kumar,Gao}, however it was also recognized that both the negative nonlocal response and specific scaling of viscous conductance with the quantum point contact slit width can arise due to various ballistic effects, see for instance Ref. \cite{Shytov}. In part for these reasons many of the current efforts are focused on direct visualization of electronic hydrodynamics. Among them are various imaging techniques \cite{Jura,Ensslin,Sulpizio,Ku,Vool,Jenkins} that have been employed to identify spatially inhomogeneous profiles of current distribution. In particular, nitrogen vacancy magnetosensors have been used to map out the current flow profile passing through constrictions with the \textit{in situ} control of parameters that govern ballistic-to-hydrodynamic crossover. In the viscous regime, the current exhibits a parabolic profile, reminiscent of the well-known Poiseuille flow. This is in contrast to the flow in the Ohmic regime, where the current is concentrated near the edge of the constrictions. In addition, geometrical control of the device including large-scale defects, wedges etc. may facilitate and amplify certain hydrodynamic features, in particular to induce vortical component of the flow when electrons are forced to move around obstacles. These ideas have also been implemented experimentally \cite{Keser,Gusev,Manfra}, perhaps with the most striking examples given by the electronic analog of the Tesla valve \cite{Smet}.    

In this work, we investigate the viscous electron flow through various geometric constrictions and calculate analytically the corresponding distributions of electric potential, current density, as well as conductivity for different boundary conditions. Additionally, analytical results are complemented by the numerical finite-element modeling for geometries that do not admit simple solutions. Primarily we are motivated by the applications of scanning tunneling potentiometry (STP) as a complementary powerful tool in obtaining the potential profile with high resolution \cite{STP}. We hope that our results can shed light on the most recent STP measurements \cite{Brar} and thus open more possibilities to identify the transport regime the electronic system is in from a combined analysis of imaging and transport data. 

The rest of the paper is organized as follows. In Sec.~\ref{sec:II} we analyze several experimentally realistic quantum point contact designs and apply the method of conformal maps to determine spatial distribution of the potential and in some cases electron flow profiles. Analytical results are also backed up by the numerical modeling. In Sec.~\ref{sec:III}, we elaborate on the physical picture of Gurzhi crossover and comment on the impact of these inhomogeneous flows on conductance across the channel. In Sec. \ref{sec:IV} we discuss correlational corrections in hydrodynamic and ballistic regimes. In particular we show the derivation of temperature dependence of momentum relaxation times arising from impurity-assisted electron scattering. In addition, we critically assess issues related to ballistic nonlocality and derive dispersive conductivity formula in a weakly impure limit. Finally, in Sec. \ref{Sec:Summary} we briefly summarize our main findings and provide perspective on interesting open questions.  

\begin{figure*}[t!]
\includegraphics[width=\linewidth]{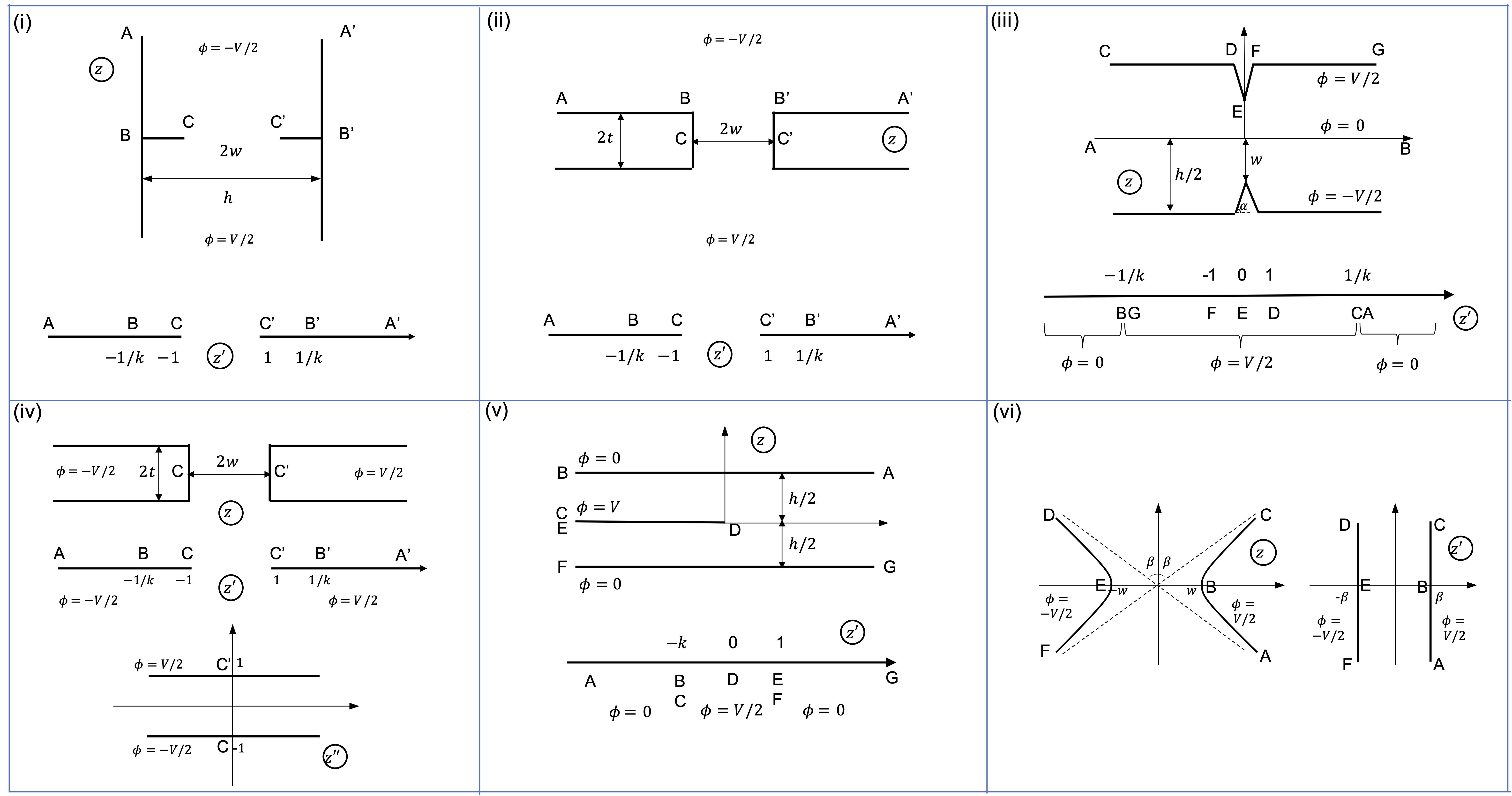}
\caption{Examples of planar constrictions for which the boundary problem of 2D hydrodynamic electronic flow can be solved exactly with the conformal mapping. (i) Flow in a finite width channel bypassing a slit. (ii) Flow through a quantum point contact with the finite width electrodes. Flow transversal to the channel trench induced between (iii) rectilinear and (iv) wedge electrodes biased by voltage $V$. (v) Flow in a capacitor type constriction with three electrodes. (vi) Flow in a smooth adiabatic point contact of hyperbolic shape.}  
\label{fig:1}
\end{figure*}

\section{Hydrodynamic Equations and Electric Potential} \label{sec:II}

Two-dimensional electronic flow in the hydrodynamic regime is described by the force balance equation \cite{Gurzhi}
\begin{equation}
	\varsigma\nabla^2\bm{u}-\frac{\bm{u}}{\tau_{\text{mr}}}=\frac{e}{m}\nabla\phi, \label{eq:force_balance}
\end{equation}
where $\bm{u}(\bm{r})$ is the hydrodynamic flow velocity, $\phi(\bm{r})$ and $n$ are the electric potential and electronic density, $\varsigma$ and $\tau_{\text{mr}}$ are the kinetic shear viscosity and the momentum relaxation time. The first term on the left hand side of Eq.~\eqref{eq:force_balance} is the viscous stress and the second term is the Ohmic dissipation. The interplay of these terms introduces the natural length scale in the problem, namely the Gurzhi length, $l_{\text{G}}\equiv\sqrt{\varsigma\tau_{\text{mr}}}$. If $l_{\text{G}} \ll w$, where $w$ is the typical size of the system such as width of the channel, the viscous stress in Eq.~\eqref{eq:force_balance} can be neglected and one is in the Ohmic (diffusive) regime. In contrast, if $l_{\text{G}} \gg w$, the Ohmic dissipation in Eq.~\eqref{eq:force_balance} is small and one is in the viscous regime. In addition, the continuity equation for the current conservation is written as
\begin{equation}
	\nabla \cdot \bm{j}=ne\nabla \cdot \bm{u}=0, \label{eq:continuity}
\end{equation}
where we assume the incompressibility of the electron fluid, i.e., the density $n$ is a constant. We examine several cases of viscous electron flow whose velocity and potential profile can be solved exactly from conformal mapping methods. Introducing the vorticity function $\omega\equiv[\nabla\times\bm{u}]_z=\partial_x v_y-\partial_y v_x$, we see that in the viscous limit, Eq.~\eqref{eq:force_balance} reduces to
\begin{equation}
	\frac{ne}{\eta}\frac{\partial \phi}{\partial x}=\frac{\partial \omega}{\partial y}, \quad \frac{ne}{\eta}\frac{\partial \phi}{\partial y}=-\frac{\partial \omega}{\partial x}, \label{eq:CR_1}
\end{equation}
where $\eta=mn\varsigma$ is the dynamic viscosity. One can observe that $-ne\phi/\eta$ and $\omega$ are the real and imaginary parts of the complex potential $\Phi$,
\begin{equation}
	\Phi(z)=-\frac{ne}{\eta}\phi+i\omega.
\end{equation}
It should be noted that knowing the vorticity function $\omega(\mathbf{r})$ still requires solving the differential equation with the proper boundary condition to restore the flow velocity field $\bm{u}(\bm{r})$. 

\subsection{Slit geometry: case I}\label{sec:II-I}

We first consider a 2D flow in an infinite strip of width $h$ obstructed by two slits with opening $2w$, see Fig.~\ref{fig:1}(i). A voltage bias $V$ is applied at the two ends of the strip. This is an example of the quantum point contact in a finite width channel. In the limit of $h\rightarrow\infty$, the problem reduces to that considered in Refs. \cite{Levitov-1,Levitov-3,Glazman}. Note that the finite width model is perhaps a more accurate representation of the experimental setting \cite{Kumar}. We intend to compute the complex potential for half of the strip geometry $ABCC'B'A'$ (the other half is a mirror symmetry). It is evident that the function 
\begin{equation}
	z'=\sin\left(\frac{\pi z}{h}\right) \label{eq:map1}
\end{equation}
maps the original geometry in the $z$ plane to the real axis (minus the opening $CC'$) in the $z'$ plane. The points $B$ and $B'$ are mapped to $z'_B=-z'_{B'}=-1$. The points $C$ and $C'$ are mapped to $z'_C=-z'_{C'}=-\sin(\pi w/h)$. The complex potential in the $z'$ plane can then be solved and is given by
\begin{equation}
	\Phi(z')=-\frac{neV}{2\eta}\frac{z'}{\sqrt{z^{'2}-z^{'2}_{C'}}}. \label{eq:Phi(z')_1}
\end{equation}
Substituting Eq.~\eqref{eq:map1} into Eq.~\eqref{eq:Phi(z')_1} and using the value of $z'_{C'}$, we obtain the complex potential in the original plane,
\begin{equation}\label{eq:Phi(z)_1}
	\Phi(z)=-\frac{neV}{2\eta}\frac{\sin(\pi z/h)}{\sqrt{\sin^2(\pi z/h)-\sin^2(\pi w/h)}}. 
\end{equation}
Setting $h\rightarrow\infty$ in Eq.~\eqref{eq:Phi(z)_1}, the complex potential reduces to that in Eq. (15) of Ref. \cite{Glazman}. The implication of such a complex potential, Eq.~\eqref{eq:Phi(z)_1}, is that the electric potential should read as
\begin{equation}
	\phi(x,y)=-\frac{\eta}{ne}\Re\Phi(z)=\frac{V}{2} \Re \frac{\sin(\pi z/h)}{\sqrt{\sin^2(\pi z/h)-\sin^2(\pi w/h)}}. \label{eq:phi(z)}
\end{equation}

\subsection{Slit geometry: case II} \label{sec:II-II}

In the second case, we consider the same geometry as in case I, except that the barrier itself also serves as electrodes on which the bias voltage $V$ is applied, see Fig.~\ref{fig:1}(iii). By symmetry, the electric potential on the midsection is identically zero. Hence, we consider the potential for the upper half of the channel, denoted by $ACDEFGB$. The mapping of the polygon $ACDEFGB$ in the $z$ plane to the upper half plane in the $z'$ plane is accomplished by Schwarz-Christoffel (SC) transformation,
\begin{equation}
	z=c\int^{z'}_0 dz' \frac{z'}{\sqrt{1-z'^2}(1-k^2z'^2)}+iw, \label{eq:map2_1}
\end{equation}
where we consider the limit that the barrier has zero width, $\alpha\rightarrow\pi/2$. The constants $k$ and $c$ are given by
\[
k=\cos\left(\frac{\pi w}{h}\right), \,\,\, c=\frac{ih}{\pi}k\sqrt{1-k^2}=\frac{ih}{2\pi}\sin\left(\frac{2\pi w}{h}\right).
\]
Evaluating the integral in Eq.~\eqref{eq:map2_1}, one obtains the mapping 
\begin{equation}
z=\frac{ih}{\pi}\arctan\left(\cot\frac{\pi w}{h}\sqrt{1-z'^2}\right), \label{eq:map2_2}
\end{equation}
or explicitly the inverse mapping
\begin{equation}
	z'=-\sgn(x)\sqrt{1+\frac{\tan^2(\pi w/h)}{\tanh^2(\pi z/h)}}.
\end{equation}
The boundary condition is mapped accordingly to the real axis of the $z'$ plane. Using the Schwarz formula for the upper half plane, the complex potential, the electric potential, and the vorticity are given by
\begin{align}
	& \Phi(z')=-\frac{neV}{2\eta}\frac{1}{\pi i} \ln\left(\frac{z'-1/k}{z'+1/k}\right), \\
	& \phi(z')=-\frac{\eta}{ne} \Re\Phi(z')=\frac{V}{2\pi}\Im\ln\left(\frac{z'-1/k}{z'+1/k}\right), \\
	& \omega(z')=\Im \Phi(z')= \frac{neV}{2\pi \eta} \Re \ln\left(\frac{z'-1/k}{z'+1/k}\right).
\end{align}
Notice that the vorticity vanishes on the segment $CG$, corresponding to the no-slip boundary condition. 

\subsection{Slit geometry: case III} \label{sec:II-III}

The next example we consider is shown in Fig.~\ref{fig:1}(ii), in which the slit has an infinite extension to the positive and negative real axes and a finite width of $2t$. Using SC transformation, the polygon $ABCC'B'A'$ in the $z$ plane is mapped to the upper half $z'$ plane by
\begin{equation}
	z=c\int^{z'}_0 dz' \sqrt{\frac{1-k^2z^{'2}}{1-z^{'2}}}=cE(\arcsin(z'),k), \label{eq:map3}
\end{equation}
where $E(\varphi,k)$ is the incomplete elliptic integral of the second kind with modulus $k \in (0,1)$. The constants $c$ and $k$ are determined by
\begin{equation}
	\frac{t}{w}=\frac{K'-E'}{E},\quad c=\frac{w}{E}, \label{eq:mapping_parameter}
\end{equation}
where $E \equiv E\left(\frac{\pi}{2},k\right)$ and $E'\equiv E\left(\frac{\pi}{2},k'\right)$ are complete elliptic integrals of the second kind.
The complex potential in the $z'$ plane is given by
\begin{equation}
	\Phi(z')=-\frac{neV}{2\eta}\frac{z'}{\sqrt{z^{'2}-1}}. \label{eq:Phi(z')_3}
\end{equation}
Since the inverse mapping $z'(z)$ is implicitly contained in Eq.~\eqref{eq:map3}, the complex potential $\Phi(z)$ in the original geometry is not cast in any simple analytical form, however we can easily consider limiting cases. In the small width limit, $t \ll w$, $B(B')$ approaches $C(C')$, therefore, $k \rightarrow 1, k' \rightarrow 0$. Using the asymptotic forms of the complete elliptic integrals,
\[
K'\approx\frac{\pi}{2}\left(1+\frac{1}{4}k'^2\right), \,\,\, E'\approx\frac{\pi}{2}\left(1-\frac{1}{4}k'^2\right), \,\,\, E \approx 1,
\]  
the constants are determined as follows,
\[
k' \approx\sqrt{\frac{4t}{\pi w}}, \,\,\, k \approx 1-\frac{1}{2}k'^2=1-\frac{2t}{\pi w}, \,\,\, c \approx w.
\]
The SC transformation \eqref{eq:map3} is then given as
\begin{equation}
	z \approx w\Big[z'+\frac{2t}{\pi w}(\arctanh z'-z')\Big],
\end{equation}
which leads to the inverse mapping,
\begin{equation}
    z' \approx \left(1+\frac{2t}{\pi w}\right)\frac{z}{w}-\frac{2t}{\pi w}\arctanh\frac{z}{w}. \label{eq:inverse}
\end{equation}
Substituting Eq.~\eqref{eq:inverse} into Eq.~\eqref{eq:Phi(z')_3}, one can find the complex potential.

In the large width limit, $t \gg w$, $k \rightarrow 0, k' \rightarrow 1$. Using the asymptotic forms of the complete elliptic integrals,
\[
K' \approx \frac{3}{2}\ln 2-\frac{1}{2}\ln(1-k'), \, E' \approx 1, \, E \approx \frac{\pi}{2}\left(1-\frac{1}{4}k^2\right),
\]  
the constants are determined as follows,
\[
k' \approx 1-\frac{8}{e^2}e^{-\pi t/w}, \, k \approx \frac{4}{e}e^{-\pi t/2w}, \, c \approx \frac{2w}{\pi}.
\]
The SC transformation \eqref{eq:map3} is then given as
\begin{equation}
	z \approx \frac{2w}{\pi} \Big[\arcsin z'-\frac{k^2}{4}\left(\arcsin z'-z'\sqrt{1-z'^2}\right)\Big].
\end{equation}
Dropping the exponentially small parameter $k$, one has the inverse mapping relation $z' \approx \sin(\pi z/2w)$, which results in the following complex potential and electric potential from Eq.~\eqref{eq:Phi(z')_3},
\begin{align}
	\Phi(z)=i\frac{neV}{2\eta}\tan\left(\frac{\pi z}{2w}\right).
\end{align}

\subsection{Slit geometry: case IV} \label{sec:II-IV}

We consider two rectilinear electrodes with bias voltage $V$ (same geometry as in Sec.~\ref{sec:II-II} but different external bias), see Fig.~\ref{fig:1}(iv). The mapping from the $z$ to $z'$ plane is computed in case IV and given by Eqs.~\eqref{eq:map3} and ~\eqref{eq:mapping_parameter}. In order to obtain the electric potential, we consider another conformal mapping function $z''=2i\arcsin(z'/\pi)$, which maps the upper half of the $z'$ plane (minus $CC'$) to the strip region $-1<y''<1,-\infty<x''<\infty$ in the $z''$-plane where the complex potential can then be easily determined,
\begin{equation}
	\Phi(z'')=i\frac{neV}{2\eta}z''.
\end{equation}

\subsection{Slit geometry: case V} \label{sec:II-V}

The next configuration is of a capacitor type, in which case a thin electrode with voltage $V$ is placed at the negative real axis $-\infty<x<0$ inside a channel of width $h$, $-h/2<y<h/2$. The edges of the channel are grounded, see Fig.~\ref{fig:1}(v). The mapping function is given as
\begin{equation}
z=\int^{z'}_{0}\!\!\!\frac{c'z'dz'}{(z'-1)(z'+k)}=c\left[\ln(1-z')+k\ln\left(1+\frac{z'}{k}\right)\right]. 
\end{equation}
The constants $k, c$ are determined as $c=h/2\pi$ and $k=h/(2\pi c)=1$. Then the required mapping
function is 
\begin{equation}
z=\frac{h}{2\pi}\ln(1-z'^2),\, z'=-\sgn(y)\sqrt{1-e^{2\pi z/h}}. 
\end{equation}
The complex potential, the electric potential, and the vorticity in the $z'$ plane are computed
from the Schwarz formula, in particular 
\begin{equation}
\Phi(z')=i\frac{neV}{2\pi\eta}\ln\left(\frac{z'-1}{z'+1}\right).
\end{equation}

\subsection{Slit geometry: case VI} \label{sec:II-VI}

The final example we consider is an adiabatic point contact that is modeled by two hyperbolic electrodes with the bias voltage $V$, see Fig. ~\ref{fig:1}(vi).
The region in the $z$ plane is bounded by
\begin{equation}
\frac{x^2}{(a\sin\beta)^2}-\frac{y^2}{(a\cos\beta)^2}=1,\quad w=a\sin\beta
\end{equation}
where $a$ is the focal length and $\beta$ is the angle that defines the asymptotes of the hyperbola. For this configuration, the function $z'=\arcsin(z/a)$ maps the region in the $z$ plane to a strip region $-\beta<x<\beta$ and $-\infty<y<\infty$ in the $z'$ plane, where the complex potential can be easily written down as $\Phi(z')=-\frac{neV}{2\eta\beta}z'$. Transforming back to the $z$ plane, we then have
\begin{equation}
\Phi(z)=-\frac{neV}{2\eta\beta}\arcsin(z/a).
\end{equation}

\begin{figure*}[t!]
\includegraphics[width=0.95\linewidth]{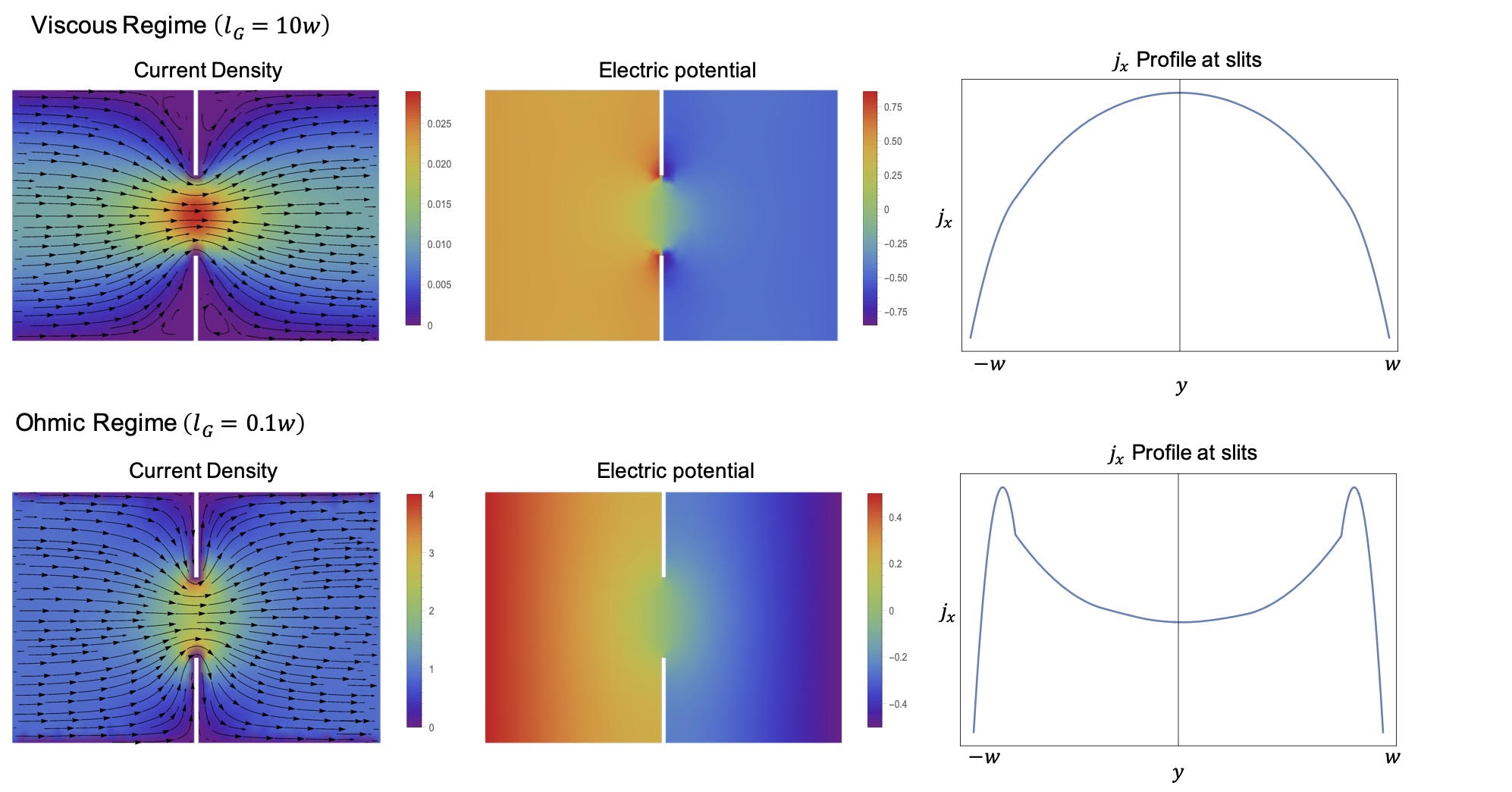}
\caption{Profiles of current density, electric potential for the flow through slits, and current distribution $j_x$ between the slits in the viscous regime
(upper panel) and Ohmic regime (lower panel). The length scale is normalized by $2w$ which is the width of the slits. In the modeling for this plot we used no-slip boundary conditions.}  
\label{fig:2}
\end{figure*}

In order to understand the qualitative behavior of potential and current distributions, as well as quantitative effects on conductance, we contrast results of conformal mapping obtained here for the hydrodynamic regime to that of Ohmic transport, see Figs. \ref{fig:2} and \ref{fig:3}.     

\section{Gurzhi Crossover} \label{sec:III}

The relationship between the lengths scales of momentum conserving collisions $l_{\text{ee}}$, momentum relaxation processes set by $l_{\text{mr}}$, and geometrical size of the constriction $w$ determines the crossover between diffusive (or ballistic) and hydrodynamic regimes. Indeed, since in the Fermi liquid theory kinematic viscosity is determined by the time scale of two-particle scattering, $\varsigma\sim v^2_F\tau_{\text{ee}}$, the Gurzhi length can be equivalently expressed as the geometrical mean $l_{\text{G}}=\sqrt{l_{\text{ee}}l_{\text{mr}}}$. This quantity strongly depends on temperature, particle density, and sample purity, and in principle can be tuned \textit{in situ} for a given device. If the channel is wide, $w\gg l_{\text{G}}$, the inhomogeneous viscous term can be dropped in the force balance equation and we recover purely Ohmic regimes with the flow velocity $\bm{u}=e\tau_{\text{mr}}\bm{E}/m$, and thus $\bm{j}=en\bm{u}=\sigma_{\text{D}}\bm{E}$ with Drude conductivity
\begin{equation} \label{sigma-D}
\sigma_{\text{D}}=\frac{e^2n\tau_{\text{mr}}}{m}.
\end{equation}       
In the opposite regime, $w\ll l_{\text{G}}$, one can neglect bulk friction and find the parabolic profile of the flow pattern in the $y$ direction:
$u(y)=\frac{d}{4}(w^2-4y^2)$, with $d=\frac{eE}{2m\varsigma}$. The spatially averaged velocity of the flow across the channel $u=w^{-1}\int^{w/2}_{-w/2}u(y)dy=dw^2/6$ gives for the current $\bm{j}=\sigma_{\text{G}}\bm{E}$ with Gurzhi conductivity 
\begin{equation}\label{sigma-G}
\sigma_{\text{G}}=\frac{e^2}{12}(nw^2)\frac{n}{\eta}. 
\end{equation}
There are several key signature features of the Gurzhi conductivity. (i) First is its quadratic scaling with the channel width that should be contrasted to the ballistic free-fermion model of Landauer transport that predicts linear in $w$ dependence. (ii) Second is the characteristic Poiseuille profile of current distribution in the uniform channel. (iii) Third is in its temperature dependence as governed by the momentum conserving collisions. As in the Fermi-liquid regime $\tau^{-1}_{\text{ee}}\propto T^2$, modulo logarithmic terms in the 2D case \cite{Novikov,Principi}, Gurzhi conductivity grows with temperature. Equivalently, the Gurzhi effect manifests itself with negative differential thermoresistivity $\partial\rho/\partial T<0$. The downturn of resistivity with increased temperature is considered as a transport  hallmark of hydrodynamic behavior in experiments \cite{LWM1,LWM2}. 

\begin{figure*}[t!]
\includegraphics[width=\linewidth]{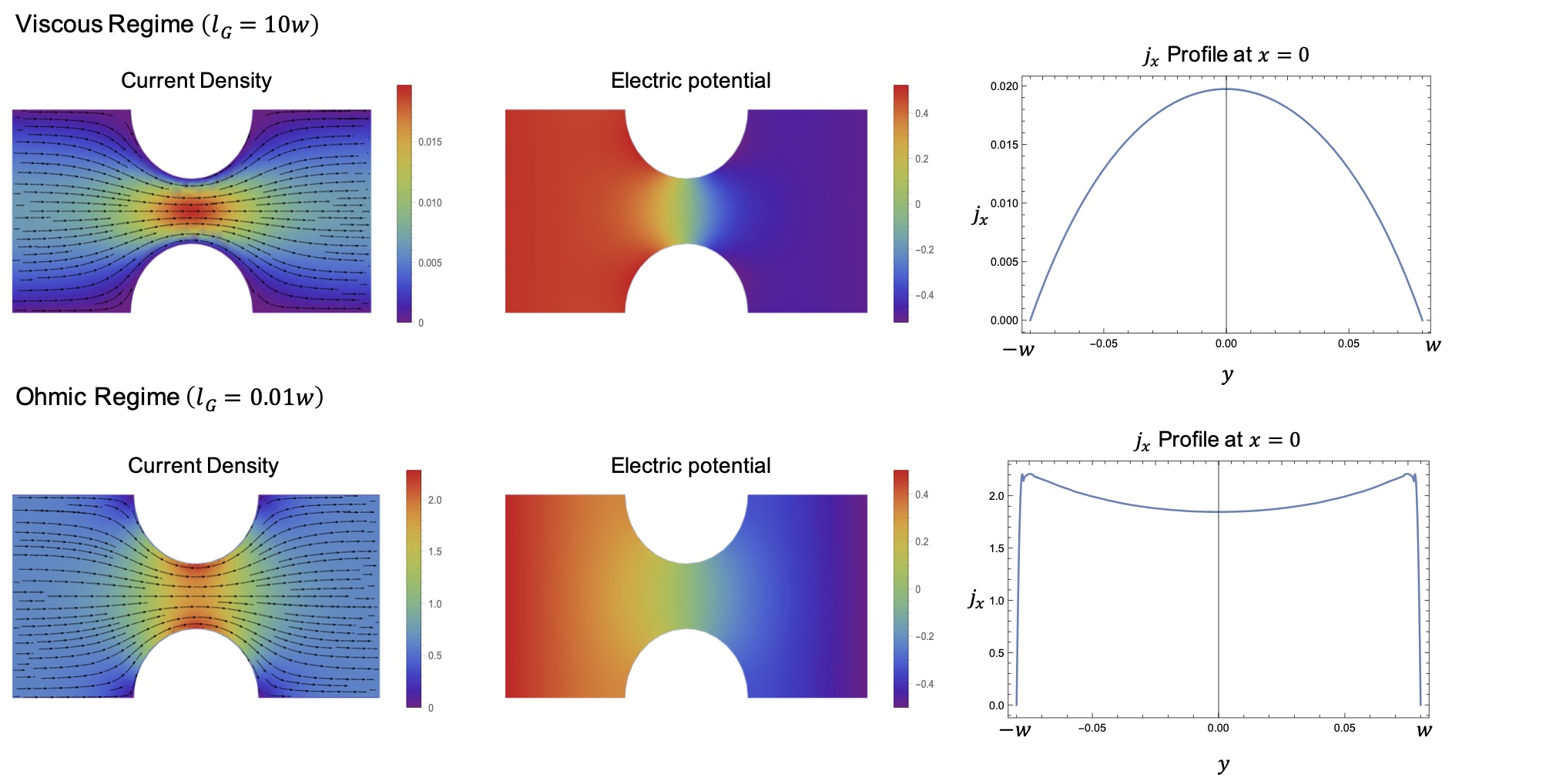}
\caption{Profiles of current density, electric potential for the flow through the semicircular constriction and $j_x$ at the narrowest
section constriction in the viscous regime (upper panel) and Ohmic regime (lower panel). As in the previous plot $2w$ is the minimum width of the
constriction.}  
\label{fig:3}
\end{figure*}

The presence of the constriction in the channel changes the profile of the flow. For comparative purposes we show in Fig. \ref{fig:2} current density and electric potential distribution for the slit geometry of Fig. \ref{fig:1}(i). The noticeable feature of these plots is convex-to-concave change in the current density profile between deep hydrodynamic and diffusive limits. It should be stressed that this feature manifests for the case of the flow with no-slip boundary conditions. In the case of no-stress boundary effects, the flow profile is concave in both cases. Boundary conditions also influence the magnitude of the Gurzhi conductivity. For instance, for the slit geometry the Gurzhi formula [Eq. \eqref{sigma-G}] has a coefficient $\pi/32$, instead of $1/12$ in the uniform channel, for the no-slip boundary, and a twice larger value in the no-stress scenario. In addition, we find that the geometrical shape of the constriction may have strong effect on the current profile. In Fig. \ref{fig:3} we plot corresponding profiles for smooth circular obstacles and notice much shallower current distribution, which in practice would be difficult to distinguish from Ohmic flow in the middle of the channel.         

\section{Correlational corrections} \label{sec:IV}

The Gurzhi-like effect, namely enhancement of conductivity in viscous flow, can occur also as the property of the system in the bulk rather than a transport phenomenon specific to a restricted geometry \cite{Hruska,Maslov,AKS,SL}. However the underlying physics is much more intricate. Gurzhi derived Eq. \eqref{eq:force_balance} from the Boltzmann equation 
\begin{equation}\label{BE}
\frac{\partial n}{\partial t}+\bm{v}\frac{\partial n}{\partial \bm{r}}+e\bm{E}\frac{\partial n}{\partial\bm{p}}
=\St_{\text{ee}}\{n\}+\St_{\text{im}}\{n\}
\end{equation}
by projecting it into the hydrodynamic mode $n(\bm{p},\bm{r},t)\to n(\varepsilon_{\bm{p}}-\bm{p}\bm{u}(\bm{r}))$, thus obtaining a macroscopic equation for $\bm{u}(\bm{r})$. The viscous term in Eq. \eqref{eq:force_balance} appears from the momentum conserving part of the collision operator $\St_{\text{ee}}\{n\}$ that captures ee-scattering processes, whereas the friction term in Eq. \eqref{eq:force_balance} comes from the electron scatterings on isolated rare impurities that are captured by $\St_{\text{im}}\{n\}$. We note that in the original work \cite{Gurzhi-KE}, Gurzhi considered the case of strongly coupled electron-phonon fluid, which is described by a specific distinct viscosity, but the essence of the derivation remains the same. Also, momentum relaxation may include other scattering pathways besides impurities such as Umklapp processes.      

The standard form of $\St_{\text{im}}\{n\}$ in Eq. \eqref{BE} is a result of a procedure in which averaging over realizations of the random impurity potential is done before solving the equation of motion \cite{Abrikosov}; as a result the effects of correlations between subsequent scattering events are lost in this treatment. Quantum interference processes renormalize Drude conductivity in the low-temperature regime. In the ballistic regime, $T\tau\gg1$, where $\tau$ is the elastic scattering time, the corresponding correction is $\delta \sigma=-\lambda\sigma_{\text{D}}(T/E_F)$, where $\lambda$ is a function of interaction parameters \cite{Zala}. The sign of the correction is not universal and depends on the details of ee-scattering. In particular, if the weak interaction is reasonably long ranged then $\lambda<0$ and the correction is positive. 

At higher temperatures correlational corrections to Drude conductivity originate from two distinct sources. The first is classical correlation effects on hydrodynamic length scales as pointed out in Ref. \cite{Hruska}, see also Sec. V in Ref. \cite{Maslov}. The underlying physics at hydrodynamic length scales, $r>l_{\text{ee}}$, basically stems from the Stokes paradox of the two-dimensional flow that gives rise to a logarithmically divergent solution to the linearized Navier-Stokes equation. The log divergence is cut by $l_{\text{ee}}$ in the lower limit and $L$ (system size or distance between impurities) in the upper limit. The second is quantum repeated back reflections from impurities on ballistic length scales, $r<l_{\text{ee}}$, as described in Ref. \cite{Levitov-2}. These processes capture multiple returns of an electron to the same impurity scatterer as a result of back reflection from ee scatterings, which is a dominant collision process in two dimensions. The back reflection has a probability $\propto\frac{1}{2}\ln(l_{\text{ee}}/a)$, where $a$ is the typical scatterer radius. Naturally, the strength of the \textit{elastic} electron-impurity scattering is reduced. The combination of the two effects gives large logarithmic correction to the Drude conductivity, 
\begin{equation}\label{sigma-visc}
\delta\sigma_{\text{Visc}}\simeq e^2(n/n_{\text{imp}})(k^2_F/\eta)\ln(L/l),
\end{equation}
which scales inversely with viscosity, where $n_{\text{imp}}$ is the scatterer concentration and $l=\sqrt{al_{\text{ee}}}$. An analog of this enhancement also appears in the system subject to long-range inhomogeneities \cite{AKS,SL}. 

The salient feature of Eq. \eqref{sigma-visc} is that the total conductivity, $\sigma=\sigma_\text{D}+\sigma_{\text{Visc}}$, manifestly violates Matthiessen's rule, according to which the resistivity is expected to be proportional to the sum of momentum relaxation rates due to various scattering processes. In contrast, an account of correlational hydrodynamic effects leads to resistivities for competing relaxation channels to add up in parallel rather than in series. 

It should be also stressed that disorder in general plays an important role in establishing hydrodynamic features in realistic samples. Indeed, in a pure system relaxation of even and odd harmonics of the electronic distribution functions occurs on a parametrically distinct timescale, $\tau_{\text{even}}/\tau_{\text{odd}}\propto (T/E_F)^2\ll1$ \cite{GKR,Ledwith}, which bears profound consequences for viscous transport in solids and for the actual form of the Gurzhi effect, see Ref. \cite{GKK}.

In addition, correlation effects can also alter the inelastic processes. The latter are described by impurity assisted electron collisions \cite{Schmid,AA,KSS}. These processes have a quantum interference origin and lead to the additional contributions to the relaxation of hydrodynamic boost velocity.  As a consequence, these terms modify the Gurzhi length scale and thus alter the temperature crossover in restricted geometries, which is relevant to numerous recent measurements.  We use the quantum kinetic equation approach to establish temperature dependence of the corresponding scattering rates in the quasiballistic parameter regime. 

\begin{figure}[t!]
\includegraphics[width=\linewidth]{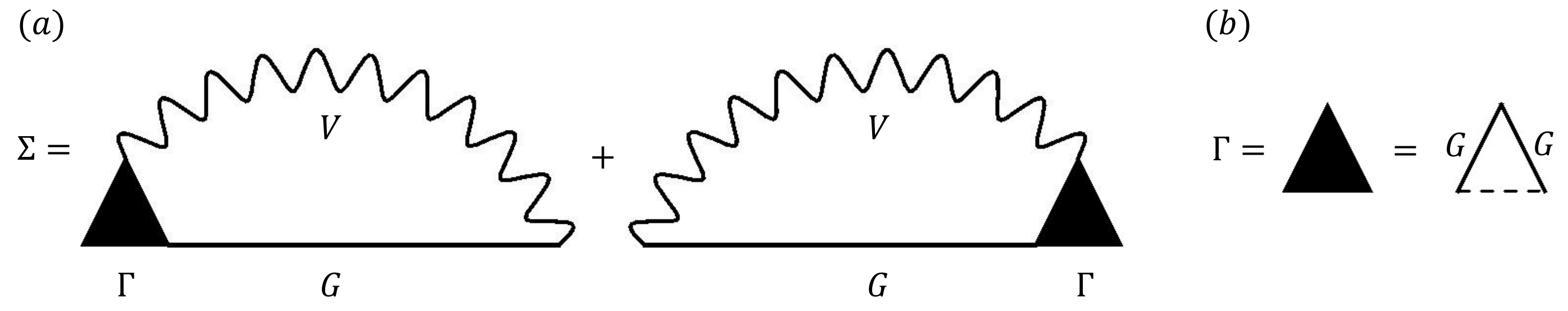}
\caption{(a) Self-energy diagram in the kinetic equation that consists of a single-particle Green's function $G$, vertex function $\Gamma$, and the interaction $V$ computed in the random-phase approximation. (b) A detailed depiction for the impurity vertex block that consists of a convolution of two Green's functions and an impurity line that brings a factor $1/(2\pi\nu\tau)$ in the model of Gaussian disorder potential. }  
\label{fig:4}
\end{figure}

\subsection{Kinetic equation}

To derive the collision integral describing impurity-assisted electron collisions, which we denote as $\St_{\text{ei}}\{n\}$ in the following, we employ the Keldysh technique \cite{Kamenev} and follow classical treatments described in Refs. \cite{KSS,Schmid,AA} adopted to the quasiballistic regime in the 2D case \cite{Zala,RW} of a weakly impure system. The collision term we are interested in,
\begin{align}
&\St_{\text{ei}}\{n\}=-\frac{i}{\pi\nu}\int\frac{d^2\bm{p}}{4\pi^2}\Im[G^R(P)]\nonumber \\ 
&\times\left\{\Sigma^K(P)-[2n(\varepsilon)-1][\Sigma^A(P)-\Sigma^R(P)]\right\}
\end{align}
is built from the retarded/advanced single-particle Green's function  
\begin{equation}
G^R(P)=[G^A(P)]^*=\frac{1}{\varepsilon-\xi_{\bm{p}}+i0},\quad \xi_{\bm{p}}=\frac{p^2-p^2_F}{2m},
\end{equation}
where $P=(\bm{p},\varepsilon)$ and $\nu$ label density of states, and $\Sigma(P)$ is the self-energy block, see Fig. \ref{fig:4} for the graphical representation. The latter includes the single-impurity vertex correction    
\begin{equation}
\Gamma^R(Q)=\frac{1}{2\pi\nu\tau}\int\frac{d^2\bm{p}}{4\pi^2}G^R(P+Q)G^A(P),
\end{equation}
where $Q=(\bm{q},\omega)$ and electron interaction resummed within the random-phase approximation
\begin{equation}
V^R(Q)=\frac{1}{V^{-1}_0(q)-\Pi^R(Q)}.
\end{equation}
Here $V_0$(q) is the bare interaction potential and the polarization operator in clean case evaluates to  
 \begin{equation}
\Pi^R(Q)=-\nu\left[1+\frac{i\omega}{\sqrt{(v_Fq)^2-\omega^2}}\right].
 \end{equation} 
 We can define the relaxation time in a usual manner as a functional derivative of the collision term with respect to the distribution function 
 \begin{equation}
 \tau^{-1}_{\text{ei}}(\varepsilon,T)=-\frac{\delta\St\{n\}}{\delta n(\varepsilon)}.
 \end{equation}
Using the standard decomposition for the Keldysh block of the self-energy in the collision kernel gives 
 \begin{align}
 \tau^{-1}_{\text{ei}}(\varepsilon,T)=-\frac{16}{\pi\nu}\int\frac{d^2\bm{p}}{(2\pi)^2}\int\frac{dQ}{(2\pi)^3}
 [N(\omega)+n(\varepsilon+\omega)]\nonumber \\ 
 \times \Im[G^R(P)]\Im[G^R(P+Q)]\Im[\Gamma^R(Q)V^R(Q)]
 \end{align}
 where $N(\omega)$ and $n(\varepsilon)$ are the bosonic and fermionic occupation functions, respectively. This result is valid under the assumptions of length and energy scales hierarchy
 \begin{equation}\label{hierarchy}
l^{-1}<q<\varkappa,\quad \tau^{-1}<\omega<v_Fq. 
 \end{equation}
Here we introduced the inverse Thomas-Fermi screening radius $\varkappa=2\pi\nu e^2$. Next we split the analysis of this scattering rate into the forward and backward scattering channels $\tau^{-1}_{\text{ei}}=  \tau^{-1}_{\text{fs}} + \tau^{-1}_{\text{bs}}$. We show that they have distinct temperature dependences. 

 \subsection{Forward scattering channel}
 
 In the $v_Fq>\omega$ domain of parameters the vertex function is real,
 \begin{equation}
 \Gamma^R(Q)=\frac{1}{\tau}\frac{1}{\sqrt{(v_Fq)^2-\omega^2}}
 \end{equation}
 and the dynamically screened interaction potential takes the form 
 \begin{equation}
 \Im V^R(Q)=-\frac{\varkappa^2}{\nu}\frac{\omega\sqrt{(v_Fq)^2-\omega^2}}{(v^2_Fq^2-\omega^2)(q+\varkappa)^2+\varkappa^2\omega^2}
 \end{equation}
 which gives us a scattering time at $\varepsilon\to0$
 \begin{equation}
 \tau^{-1}_{\text{fs}}=\frac{4\varkappa^2}{\pi^2\tau\nu}\int^{\infty}_{0}d\omega \omega[N(\omega)+n(\omega)]K_{\text{fs}}(\omega).
 \end{equation}
In the resulting kernel,
\begin{equation}
K_{\text{fs}}=\int^{\infty}_{\frac{\omega}{v_F}}\frac{qdq}{\sqrt{(v_Fq)-\omega^2}}\frac{1}{((v_Fq)^2-\omega^2)(q+\varkappa)^2+\varkappa^2\omega^2}
\end{equation}
it is sufficient to approximate $(q+\varkappa)^2\to \varkappa^2$ because of the constrain on relevant momenta formulated in Eq. \eqref{hierarchy}. 
The remaining integral can be done simply by scaling and change of variable, so that we find 
 \begin{equation}
 K_{\text{fs}}(\omega)=\frac{\pi}{2}\frac{1}{\varkappa^2v^2_F|\omega|}.
 \end{equation}
The numerical coefficient comes from the integral $\quad \int^{\infty}_{1}\frac{dx}{x\sqrt{x^2-1}}=\frac{\pi}{2}$. The remaining frequency integral is log divergent. The ultraviolet limit should be cut at $E_F$, whereas the infrared limit by $\tau^{-1}$ so that the regularized expression is 
 $\int^{E_F}_{\tau^{-1}}d\omega [N(\omega)+n(\omega)]\simeq T\ln(E_F\tau)$. The most important part comes from $N(\omega)$, whereas $n(\omega)$ gives a factor of $\ln(2)$, however retaining this term exceeds the accuracy of our logarithmic approximation.  
As a result,
\begin{equation}
\tau^{-1}_{\text{fs}}=\frac{4T}{E_F\tau}\ln(E_F\tau)
\end{equation}
the relaxation rate is linear in temperature. 

\subsection{Backscattering channel}
 
To address the contribution of backscattering we need a proper generalized expression for the polarization operator near momenta $k=q-2k_F$ with $k\ll2k_F$. For this purpose one needs to retain $q^2/2m$ in the Green's function when computing $\Pi^R(Q)$. Carrying out this analysis we find a modified form of the corresponding collision kernel in the kinetic equation 
\begin{align}
K_{\text{bs}}(\omega)&=\int^{\infty}_{\frac{\omega}{v_F}}\frac{2k_Fdk}{\sqrt{8E_F(v_Fk+\omega)}}\nonumber \\ 
&\times \frac{1}{8E_F(v_Fk+\omega)(2k_F+\varkappa)^2+\omega^2\varkappa}
\end{align} 
The last term in the denominator $\omega^2\varkappa$ can be neglected as compared to the first term. The resulting momentum integration then gives 
\begin{equation}
K_{\text{bs}}(\omega)=\frac{2\sqrt{2}k_F}{(8E_F)^{3/2}(2k_F+\varkappa)^2}\frac{1}{v_F\sqrt{|\omega|}}.
\end{equation} 
The final result for the backscattering rate is then
\begin{equation}
\tau^{-1}_{\text{bs}}=\frac{\zeta[3/2](2\sqrt{2}-1)}{2\sqrt{2\pi}}\tau^{-1}\left(\frac{\varkappa}{2k_F+\varkappa}\right)^2\left(\frac{T}{E_F}\right)^{3/2},
\end{equation}
where $\zeta[x]$ is the Riemann zeta function. We conclude that forward scattering dominates. 

 \subsection{Remarks on non-locality}
 
The purpose of this section is to simply remind that nonlocal effects and inhomogeneous distribution of currents can naturally occur in 
impure conductors, so that this aspect of conduction properties is obviously not unique to the hydrodynamic regime of viscosity dominated flow. Indeed, within the linear response theory, one defines a nonlocal conductivity tensor $\sigma_{\alpha\beta}(\bm{r},\bm{r}')$ as an integral relationship between current at a point $\bm{r}$ and the local electric field at a point $\bm{r}'$ as follows 
\begin{equation}
j_{\alpha}(\bm{r})=\int d\bm{r}'\sigma_{\alpha\beta}(\bm{r},\bm{r}')E_{\beta}(\bm{r}')
\end{equation} 
In general, $\sigma_{\alpha\beta}(\bm{r},\bm{r}')$ receives short range and long range contributions. The Kubo formula gives practical tools to express nonlocal conductivity in terms of the electronic Green's function, which can be written in the form \cite{Kane-Lee} 
\begin{align}\label{sigma-rr}
\sigma_{\alpha\beta}(\bm{r},\bm{r}')=-\frac{e^2}{\pi m^2}\left[\partial_\alpha \Im G^R_\epsilon(\bm{r}, \bm{r}')\partial'_\beta \Im G^R_\epsilon(\bm{r}', \bm{r})\right. \nonumber \\ 
\left.-\Im G^R_\epsilon(\bm{r}, \bm{r}') \partial_\alpha \partial'_\beta \Im G^R_\epsilon(\bm{r}', \bm{r}) \right],
\end{align}
where $\partial_\alpha \equiv \partial/\partial r_\alpha, \, \partial'_\alpha \equiv \partial/\partial r'_\alpha$. Our intent is to calculate nonlocal conductivity in the simplest model of a disordered 2D conductor in the quasiballistic regime with isotropic scatterers. We use the impurity diagram technique to find the disorder averaged single-particle Green's function  
 \begin{equation}\label{ImG}
 \Im G(\bm{r},\bm{r}')=-\pi \nu J_0(k_FR)e^{-R/2l},
 \end{equation}
where $R \equiv |\bm{R}|=|\bm{r}-\bm{r}'|$, $l=v_F\tau$ is the disorder mean free path, and $J_n(x)$ is the Bessel function of the first kind. Inserting Eq. \eqref{ImG} into Eq. \eqref{sigma-rr}, performing differentiation, and retaining terms containing the highest power in $k_Fl\gg1$, we find  
\begin{align}
\sigma_{\alpha\beta}(\bm{r},\bm{r}'&)=\frac{e^2(\pi\nu)^2}{\pi m^2}\frac{e^{-R/l}R_\alpha R_\beta}{(Rl)^2}\nonumber \\ 
&\times (k_Fl)^2\left[J^2_1(k_FR)+J_0(k_FR)J'_1(k_FR)\right].
\end{align}
At this point it is convenient to introduce the Fourier transform 
\begin{align}
&\sigma_{\alpha\beta}(\bm{q})=\int d^2R\, e^{i\bm{q} \cdot \bm{R}}\sigma_{\alpha\beta}(\bm{r},\bm{r}')=\delta_{\alpha\beta}\frac{e^2}{4}(k_Fl)^2\nonumber \\ & \times\int^\infty_0 dx e^{-x}x J_0(qlx)\left[J^2_1(k_Flx)+J_0(k_Flx)J'_1(k_Flx)\right]
\end{align}
where we introduced the dimensionless variable $x=R/l$. To complete the remaining integral we again take advantage of the large parameter $k_Fl\gg1$ and
thus utilize the asymptotic form of the Bessel function at large arguments: $J_n(x) \sim \sqrt{\frac{2}{\pi x}}\cos\left(x-\frac{n\pi}{2}-\frac{\pi}{4}\right)$  and $J'_n(x) \sim -\sqrt{\frac{2}{\pi x}}\sin\left(x-\frac{n\pi}{2}-\frac{\pi}{4}\right)$. As a result we find 
\begin{align}\label{sigma-ql}
\sigma_{\alpha\beta}(\bm{q}) = \delta_{\alpha\beta}\sigma_{\text{D}} \int^\infty_0 dx e^{-x} J_0(qlx)= \frac{\delta_{\alpha\beta}\sigma_{\text{D}}}{\sqrt{1+(ql)^2}}, 
\end{align}
where $\sigma_{\text{D}}=\frac{e^2}{2\pi}(k_Fl)$ is the Drude conductivity in 2D. Clearly the inverse Fourier transform to the real space coordinates
will result in a highly nonlocal kernel, $\propto e^{-|\bm{r}-\bm{r}'|/l}/|\bm{r}-\bm{r}'|$, between the electric field and the current. This ultimately renders an inhomogeneous current distribution in the sample whose actual profile depends on the geometry of the device as well as specular or diffusive boundary scattering. We note that quantum aspects of nonlocal transport from weak electron-impurity scattering were discussed recently in Ref. \cite{Hui}. The result for conductivity at finite wave vector $\sigma(\bm{q})$ was derived perturbatively from the Kubo formula via small-$q$ expansion to the leading $q^2$ order that thus generates spatial inhomogeneity, $q^2\to -\nabla^2$, appropriate for the diffusive regime $ql<1$. 
The result captured by Eq. \eqref{sigma-ql} gives a proper generalization for the quasiballistic regime $ql>1$.  
 
  \section{Summary and perspective}\label{Sec:Summary}
  
In this work we considered hydrodynamic aspects of electronic transport in mesoscopic structures including quantum point contacts. The focus of the work was on effects of device geometry on viscous profiles of inhomogeneous current distributions, impact of impurity scattering, and boundary conditions across the parameter range of the Gurzhi effect that describe crossover from the Ohmic to Stokes transport regime. We used the method of conformal transformation to map out spatial distribution of electric potential that can be measured by the STP method. 

We also derived a kinetic equation that captures quantum correlations stemming from the impurity assisted quasiballistic interference effects that lead to renormalization of momentum-relaxing scattering due to electron collisions $\tau^{-1}_{\text{mr}}=\tau^{-1}\left[1+\frac{4T}{E_F}\ln(E_F\tau)\right]$. The correction could become quite noticeable at the elevated temperatures at the onset of the hydrodynamic regime. This effectively changes temperature dependence of the Gurzhi length and thus crossover in the restricted geometries. This also complicates extraction of the temperature dependence of electron viscosity from transport experiments. These effects will be relevant also in the nonhydrodynamic regime where nonlocality plays an important role. The mechanism presented here would then yield a $T$-dependent response that could be falsely interpreted as viscous effects.  
 
Although hydrodynamic definition of viscosity assumes clean systems, it can also be introduced in the context of disordered conductors as a linear response that characterizes a change of the stress tensor under time-dependent deformations \cite{Bradlyn}. As such, the nonlocal effects in disordered systems open interesting perspectives to further study the relation between Hall viscosity, second derivative of Hall conductivity with respect to the momentum, Wen-Zee shift, and interaction crossover from classical to quantum hydrodynamic regimes \cite{Hoyos,Scaffidi,Burmisrov}. We are certain these topics deserve further attention that we leave for future work.    

\textit{Note added}: Recently, we became aware of a related study \cite{Tikhonov}, where electron flow in the hydrodynamic regime was studied with an account of heating effects due to inhomogeneous electron temperature profiles. We computed the distribution of heat production from a viscous flow and obtained a similar symmetric Landauer-dipole feature [e.g., their Fig. 1(d)]. We confirm that the asymmetric feature and the suppression of Landauer dipole can be reproduced in the extended framework with an additional equation for temperature profile $T(\bm{r})$. These additional aspects of the problem may be important in the interpretation of experimental findings reported in Ref. \cite{Brar}. 
 
\section*{Acknowledgments}

We appreciate discussions with Wyatt Behn, Victor Brar, and Zachary Krebs that motivated this study. We thank Aaron Hui and Eun-Ah Kim for communication regarding Ref. \cite{Hui} 
and Igor Gornyi for bringing Ref. \cite{Tikhonov} to our attention. We are grateful to Igor Burmistrov for discussions on the diagrammatic evaluation of the localization corrections to the conductivity at finite momentum, to Boris Narozhny for discussion of boundary conditions in the viscous regime, and to Shahal Ilani for discussions on the most recent imaging measurements of electronic flow in various geometries. This work was supported by the U.S. Department of Energy, Office of Science, Basic Energy Sciences Program for Materials and Chemistry Research in Quantum Information Science under Award No. DE-SC0020313. M.K. acknowledges financial support from the Israel Science Foundation, Grant No. 2665/20 and in part by BSF Grant No. 2016317. This paper was finalized during the workshop ``From Chaos to Hydrodynamics in Quantum Matter" at the Aspen Center for Physics, which is supported by National Science Foundation Grant No.PHY-1607611.

\end{document}